\documentclass[11pt]{article}
\usepackage{moriond,epsfig}
\usepackage{wrapfig}
\usepackage{subfigure}

\def\be{\begin{equation}}
\def\ee{\end{equation}}
\def\bea{\begin{eqnarray}}
\def\eea{\end{eqnarray}}

\begin{document}
\vspace*{4cm}
\title{MEASUREMENTS OF STANDARD MODEL PROCESSES AT ATLAS}

\author{ J. ALISON~\footnote{on behalf of the ATLAS Collaboration} }

\address{Department of Physics, University of Pennsylvania\\
  209 South 33rd st.  Philadelphia Pa., United States}

\def\Wg{\textit{W$\gamma$} }
\def\Zg{\textit{Z$\gamma$} }
\def\WW{\textit{WW} }
\def\WWZ{\textit{WWZ} }
\def\dspde{\frac{\mathrm{d}\sigma_{W^+_{\mu}}}{\mathrm{d}\eta_{\mu}}}
\def\dsnde{\frac{\mathrm{d}\sigma_{W^-_{\mu}}}{\mathrm{d}\eta_{\mu}}}
\def\pt{$p_T$}
\def\et{$E_T$}
\def\MeT{$E_T^{miss}$}
\def\MeTRel{$E_T^{miss,Rel}$ }
\def\mT{$m_T$ }
\def\W{\textit{W} boson }
\def\Wjet{\textit{W}+jet }
\def\Ws{\textit{W}'s }
\def\Wt{\textit{Wt}}
\def\Z{\textit{Z} boson }
\def\pp{\textit{pp} }
\def\t{\textit{t}}
\def\u{\textit{u}}
\def\d{\textit{d}}
\def\s{\textit{s}}
\def\b{\textit{b}}

\maketitle\abstracts{
With over \mbox{45 pb$^{-1}$} of 7 TeV \pp collisions recorded, the ATLAS Standard Model physics program is well under way.
These proceedings survey the latest tests of the Standard Model at this unprecedented energy scale.
An overview of recent ATLAS results is given. 
Measurements of the \W charge asymmetry, di--boson production, and single top--quark production are highlighted. 
}

\section{Introduction}

Measurements of Standard Model (SM) processes have been the flagship of the ATLAS~\cite{ref:Atlas} physics program in 2010.
These measurements cover a wide range of topics from soft QCD measurements of particle multiplicities and the total \pp inelastic cross section, through QCD measurements of inclusive jet production, photon production, and top-quark pair production, to electro-weak measurements of vector--boson properties, di--boson production, and single top--quark production.

In addition to being a rich source of physics, Standard Model processes serve as standard candles from which the detector performance can be understood. 
The expected SM signals can be used to commission the detector and refine analysis techniques in preparation for the unexpected.
The physics objects used in SM measurements - charged leptons, missing transverse energy, photons, and jets - are critical for all physics analyses.
The understanding of these objects, gained initially through SM measurements, is of wide--ranging importance for all the physics done at ATLAS. 

\begin{figure}[ht]
\begin{center}
\includegraphics[width=14cm,height=6cm]{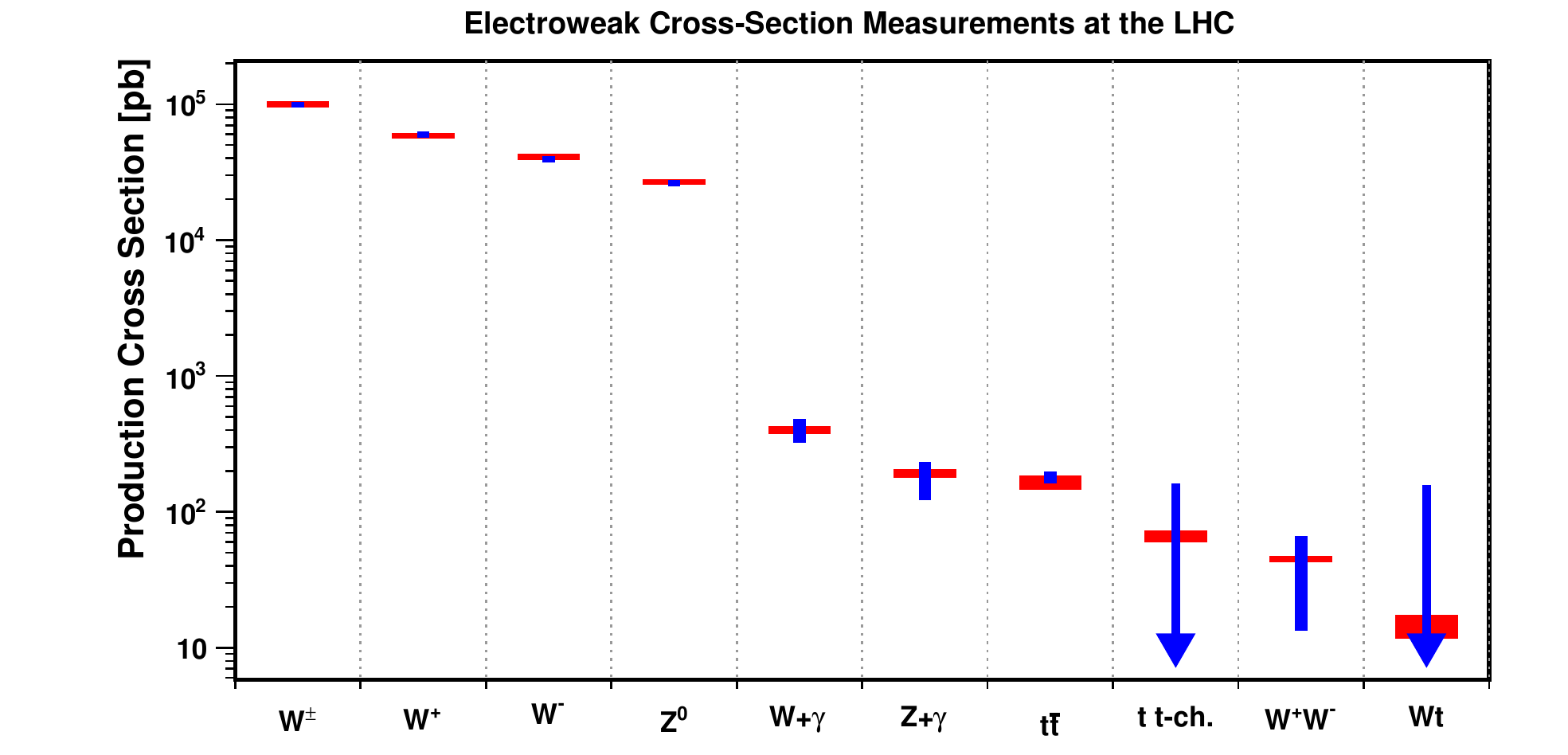}\label{fig:xSecs}
\caption{Electro--weak production cross sections at the LHC.  Theoretical predictions are shown in red, the corresponding ATLAS measurements are given in blue. See references [2-6] for cross section measurements.} 
\end{center}
\end{figure}

These proceedings will focus on recent electro--weak measurements. 
The measurement of the \W charge asymmetry is presented in Section~\ref{sec:WAsymm}, followed by the measurement of the \Wg and \Zg cross sections in Section~\ref{sec:WgZg}.
Section~\ref{sec:WW} presents the measurement of the \WW production cross section and results on single top--quark production are given in Section~\ref{sec:sTop}.
As can be seen in Figure~\ref{fig:xSecs}, the electro--weak measurements presented in these proceedings span several orders of magnitude in production cross section.
The varying amounts of signal and sources of background across this broad spectrum pose unique challenges to the different analyses presented here.  

\section{\W Charge Asymmetry}\label{sec:WAsymm}

The \W charge asymmetry is particularly interesting because it is sensitive to the parton distributions functions, PDFs, of the proton.
A precision measurement of the asymmetry can be used to constrain the PDFs of the valance quarks in the $10^{-3} - 10^{-4}$ region of momentum fraction.~\cite{ref:WAsymmCONF}

The \W charge asymmetry has been measured by ATLAS in the muon decay channel as a function of the muon pseudo--rapidity, $\eta_{\mu}$.~\cite{ref:WAsymmCONF}
The asymmetry, defined as
\begin{equation}
        A_{\mu} = \frac{\dspde - \dsnde}{\dspde + \dsnde},
\end{equation}
consists of the ratio of production cross sections, which has the advantage that many of the experimental uncertainties cancel.
The measurement was performed by selecting events containing reconstructed muons with transverse momentum, \pt, above 20 GeV, and missing transverse energy, \MeT, above 25 GeV.
The events were additionally required to contain a reconstructed \W candidate with a transverse mass, \mT of more than 40 GeV.  
This selection led to $1.3 \times 10^5$ \W candidates, with an estimated background of seven percent predominantly from background from other electro--weak processes.                                                         

\begin{figure}
  \begin{center}
    \subfigure[]
    {\includegraphics[width=7cm,height=6.3cm]{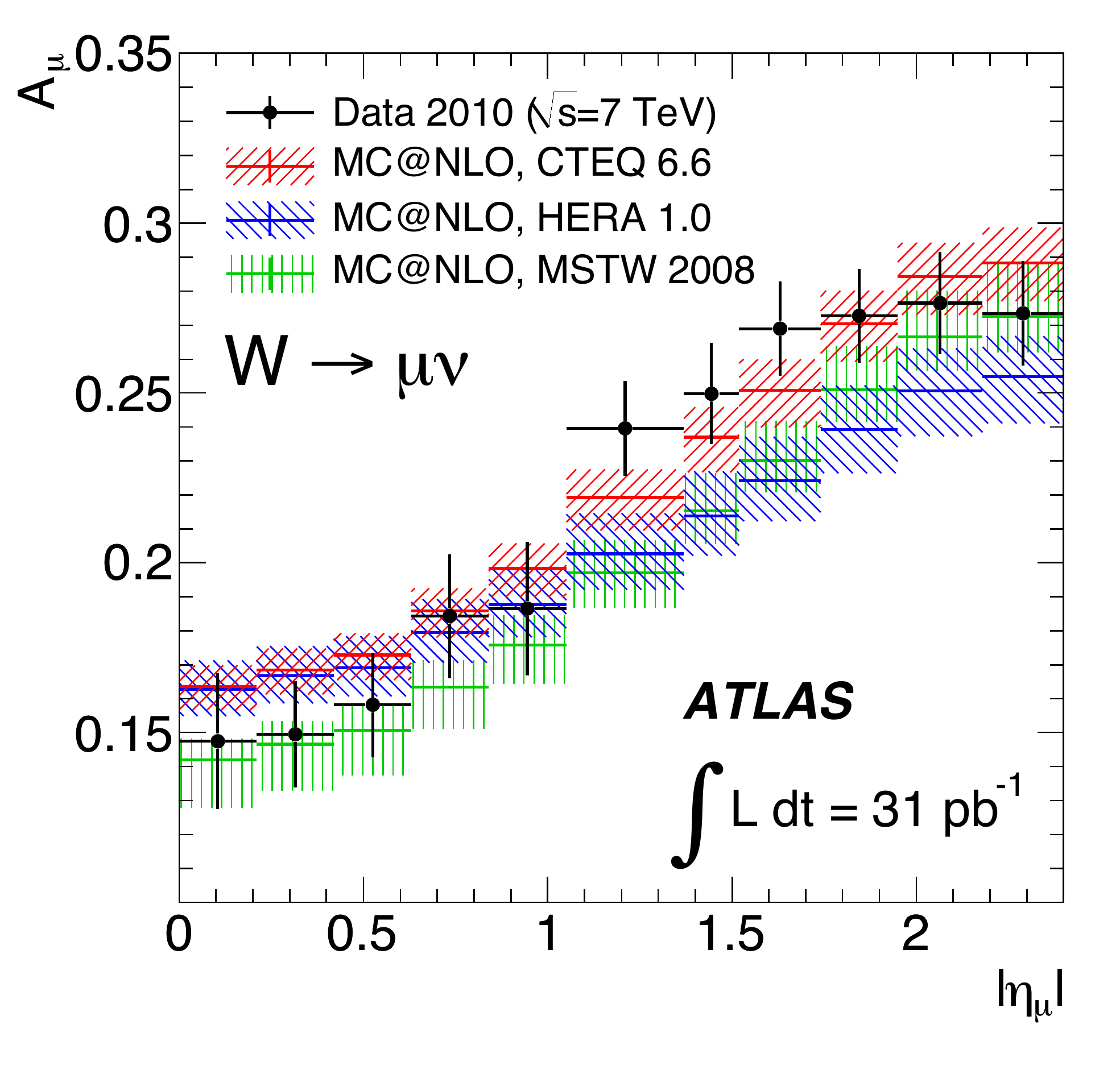}\label{fig:wAsymm}}
    \subfigure[]
    {\includegraphics[width=7cm,height=6cm]{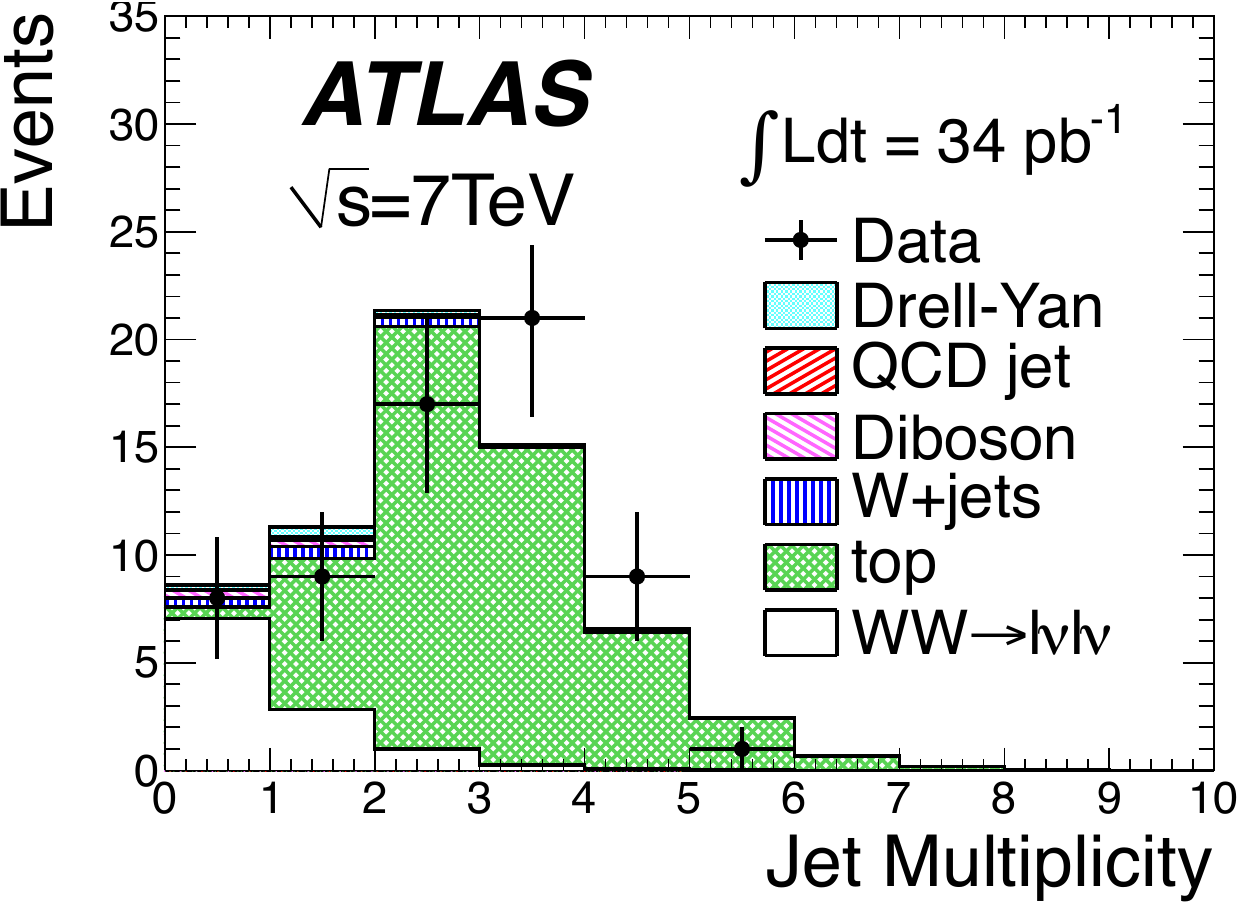}\label{fig:WWN}}
    \caption{\subref{fig:wAsymm} The \W asymmetry as a function of muon pseudo--rapidity.\protect \cite{ref:WAsymmCONF} The measurement is shown in black along with theoretical predictions from several parametrizations of the parton distribution function. \subref{fig:WWN} The jet multiplicity distribution for events satisfying the di--lepton plus \MeT\ selection in the \WW cross section analysis.\protect \cite{ref:WWCONF} The zero--jet bin is used for the signal extraction. }
    \label{fig:number2}
  \end{center}
\end{figure}

The measured \W asymmetry is shown as a function muon pseudo--rapidity in Figure~\ref{fig:wAsymm}.
The asymmetry rises with $\eta_{\mu}$, as predicted by theory, with statistical and systematic uncertainties that are comparable in each $\eta_{\mu}$-bin.
The limiting systematic uncertainties come from uncertainties on the trigger and muon identification efficiencies, which vary across $\eta_{\mu}$ from 1--7\%.
These systematics are currently limited by statistics in control regions and will decrease with further data taking.

The predicted \W asymmetry from several global fits to the proton PDF are also shown in Figure~\ref{fig:wAsymm}.
The current experimental uncertainty of the ATLAS measurement is already comparable to those of the global fits.
Future measurements of the \W asymmetry will constrain the proton PDFs.

\section{\Wg and \Zg Cross Sections}\label{sec:WgZg}


ATLAS has performed the measurement of the \Wg and \Zg cross sections in the leptonic decay channels of the \textit{W} and \textit{Z} bosons~\cite{ref:WgZgCONF}
\footnote{Here, and in the rest of these proceedings, the leptonic decay channels only refer to the decays of \textit{W} or \textit{Z} bosons to electrons or muons, including the decays to electrons and muons through $\tau$s. Decays to hadronicaly decaying $\tau$s have been neglected.}.
The measurement of these \Wg and \Zg processes provides a test of the electro--weak model. 
A photon can be produced in association with a \W or a \Z through the initial state radiation (ISR) of a photon off of an incoming quark, or by the final state radiation (FSR) of a photon off of the \W or \Z decay products.   
The \Wg process can also be produced through an additional diagram in which the photon is directly radiated from the \textit{W} boson.
This diagram is sensitive to the triple gauge coupling (TGC) predicted by the Standard Model. 

Events were selected containing a \textit{W} or \Z candidate and an isolated reconstructed photon with transverse energy, \et, greater than 15 GeV.
The \W candidates were required to have an electron or muon with \et\ greater than 20 GeV, \MeT\ greater than 25 GeV, and \mT above 40 GeV.
\Z candidates were required to have two electrons or muons, each with \et\ greater than 20 GeV, and having an invariant mass above 40 GeV.
The reconstructed photons were required to be well isolated and to be separated from the reconstructed lepton by more than 0.7 in $\Delta R$.\footnote{$\Delta R^2$ is defined as $\Delta R^2 = {\Delta \phi}^{2} + {\Delta \eta}^2$, where $\Delta \phi (\Delta \eta)$ is the difference in $\phi (\eta)$ between the photon and the lepton. } 
The event yields and estimated backgrounds are given in Tables~\ref{tab:Wg} and~\ref{tab:Zg}.
The \Wjet background in the \Wg analysis is derived from control regions in the data.

\begin{table}[htb]
  \begin{center}
    \begin{tabular}{c|c|c}
      \Wg   & electron-channel & muon-channel   \\
      \hline
      Event Yield & 95 & 97 \\
      \hline
      \W + jet Bkg  & 16.9 $\pm$ 5.3 $\pm$ 7.3 & 16.9 $\pm$ 5.3 $\pm$ 7.4 \\ 
      \hline
      EW Bkg  & 10.3 $\pm$ 0.9 $\pm$ 0.7 & 11.9 $\pm$ 0.8 $\pm$ 0.8 \\ 
      \hline
    \end{tabular}
   \caption{\Wg yields and background estimates. The \Wjet background is derived from control regions in the data, whereas the remaining electro--weak background is taken from simulation.}
    \label{tab:Wg}
  \end{center}
\end{table}

\begin{table}[htb]
  \begin{center}
    \begin{tabular}{c|c|c}
      \Zg   & electron-channel & muon-channel  \\
      \hline
      Event Yield & 25 & 23 \\
      \hline
      Background  & 3.7 $\pm$ 3.7 & 3.3 $\pm$ 3.3\\
      \hline
    \end{tabular}
   \caption{\Zg yields and background estimates. The background estimates are taken from simulation.}
    \label{tab:Zg}
  \end{center}
\end{table}

The results of the \Wg and \Zg cross section measurements are presented in Tables~\ref{tab:WgResults} and~\ref{tab:ZgResults}.
The limiting systematic uncertainties are from uncertainties associated to the photon reconstruction and identification, uncertainties on the background predictions, and uncertainties associated with the signal acceptance.
The measured cross sections are in good agreement with the NLO SM prediction.
Future measurements of the \Wg and \Zg processes will constrain new physics in anomalous TGCs.
\begin{table}[htb]
  \begin{center}
    \begin{tabular}{c|c}
      \Wg   & Cross Section [pb] \\
      \hline
      e-channel & 48.9 $\pm$ 6.6 (stat) $\pm$ 8.3 (sys) $\pm$ 1.7 (lumi)\\
      \hline
      $\mu$-channel & 38.7 $\pm$ 5.3 (stat) $\pm$ 6.4 (sys) $\pm$ 1.3 (lumi)\\
      \hline
      \hline
      SM NLO Prediction & 42.1 $\pm$ 2.7 (sys) \\
      \hline
    \end{tabular}
   \caption{The measured \Wg cross sections in electron and muon channel compared to the SM NLO predictions.  The cross sections are reported for $E_T^{\gamma} > 15$ GeV and $\Delta R(l,\gamma) > 0.7$.}
    \label{tab:WgResults}
  \end{center}
\end{table}

\begin{table}[htb]
  \begin{center}
    \begin{tabular}{c|c}
      \Zg   & Cross Section [pb] \\
      \hline
      e-channel & 9.0 $\pm$ 2.5 (stat) $\pm$ 2.1 (sys) $\pm$ 0.3 (lumi)\\
      \hline
      $\mu$-channel & 5.6 $\pm$ 1.4 (stat) $\pm$ 1.2 (sys) $\pm$ 0.2 (lumi)\\
      \hline
      \hline
      SM NLO Prediction & 6.9 $\pm$ 0.5 (sys) \\
      \hline
    \end{tabular}
   \caption{The measured \Zg cross sections in electron and muon channel compared to the SM NLO predictions.  The cross sections are reported for $E_T^{\gamma} > 15$ GeV and $\Delta R(l,\gamma) > 0.7$.}
    \label{tab:ZgResults}
  \end{center}
\end{table}

\section{\WW Cross Section}\label{sec:WW}
Similar to the \Wg and \Zg processes, another process which tests the electro--weak model is \WW di--boson production.
The \WW final state is produced primarily through quark annihilation at the LHC, and includes a diagram sensitive to the \WWZ TGC predicted by the SM.  
In addition to being sensitive to new physics through anomalous TGCs, the \WW process is also important because it is the dominant background to searches for the Higgs boson in which the Higgs decays to pairs of \textit{W} bosons.

ATLAS has performed the \WW cross section measurement in the fully leptonic decay channels of the \Ws.~\cite{ref:WWCONF}
The event signature is two high--\pt\ isolated leptons with large missing energy. 
The jet multiplicity distribution of events satisfying the di--lepton plus \MeT\ selection
\footnote{The \MeT\ requirement is on the \MeT\ relative to the nearest lepton, \MeTRel.  \MeTRel is defined as \MeTRel $= sin(\phi) \times $\MeT, if $\phi < \frac{\pi}{2}$, otherwise \MeTRel $=$ \MeT. $\phi$  is the angle between the missing energy and the nearest lepton. Events are required to have greater than 40 GeV relative \MeT\ in the \textit{ee} and $\mu\mu$ channels, and greater than 25 GeV in the $e\mu$ channel.} 
 is shown in Figure~\ref{fig:WWN}.

The large remaining background from top--quark production is reduced by requiring that the event contain no reconstructed high \pt\ jets within the ATLAS acceptance. 
Eight signal candidates pass the full selection, one in the \textit{ee}-channel, two in the $\mu\mu$-channel, and five in the $e\mu$-channel. 
The background estimation is provided in Table~\ref{tab:WWBkg}.  
The  \Wjet estimate is made from control regions in the data, whereas the remaining electro--weak backgrounds are taken from simulation and cross checked with data--driven procedures.

\begin{table}[htb]
  \begin{center}
    \begin{tabular}{c|c}
      Background   & Events  \\
      \hline
      Drell-Yan & 0.23 $\pm$ 0.15 (stat) $\pm$ 0.17 (sys) \\
      \hline
      Top Quark & 0.53 $\pm$ 0.12 (stat) $\pm$ 0.28 (sys) \\
      \hline
      \Wjet & 0.54 $\pm$ 0.32 (stat) $\pm$ 0.21 (sys) \\
      \hline
      Other Di-boson & 0.38 $\pm$ 0.04 (stat) $\pm$ 0.04 (sys) \\
      \hline
      \hline
      Total Bkg. & 1.68 $\pm$ 0.37 (stat) $\pm$ 0.42 (sys) \\
      \hline
    \end{tabular}
   \caption{The estimated background for the \WW cross section measurement. The background from \Wjet was estimated from control regions in data, whereas the other background as estimated from simulation and cross checked by data-driven methods.}
    \label{tab:WWBkg}
  \end{center}
\end{table}

The measured \WW cross section is $41^{+20}_{-16} (stat.) \pm 5 (syst.) \pm 1 (lumi.) $ pb, which is to be compared to the SM NLO prediction of \mbox{$44 \pm 3$ pb}. 
The dominant uncertainty on the cross section measurement is the statistical uncertainty on the number signal events, 44\%.  
The systematic uncertainty is 16\%, and is dominated by uncertainties on the background modeling and signal acceptance.
Further studies of the \WW process will constrain new physics through measurements of anomalous TGCs and will be critical for understanding the background in the search for the Higgs boson. 

\section{Single Top-Quark Production}\label{sec:sTop}

Single top-quark production is a direct probe of the CKM element $V_{tb}$.
A precise measurement of the single top-quark cross section will provide a determination of $V_{tb}$ without relying on unitarity constraints.
Single top-quark production proceeds through three modes, each with a distinct experimental signature.
The \t-channel production has the largest expected contribution, $\sim$\mbox{65 pb}, and leads to a top quark and either a \u\ or \d-quark in the final state. 
\Wt-production is expected to have the second largest contribution, $\sim$\mbox{15 pb}, and has a top-quark and a \W in the final state.
The \s-channel is the smallest expected single top-quark contribution at the LHC, $\sim$\mbox{4 pb}, and produces a top quark and bottom quark in the final state.  
Each of the individual single top-quark production modes is sensitive to different forms of new physics.~\cite{ref:singleTopBSM}
With the 2010 data set, ATLAS has performed searches for the single top-quark in both the \t-channel and \Wt-production modes.~\cite{ref:singleTopCONF}

The \t-channel single top-quark analysis has been performed in the leptonic decay mode of the top-quark.
Events were selected with one high-\pt\ electron or muon, large \MeT, and two jets, one of which was identified as a \b-quark. 
The \mT of the lepton and \MeT\ system was required to be consistent with coming from a \W and the reconstructed top-quark mass was required to be between 130 and 210 GeV.
To enhance sensitivity, the analysis was performed separately in the positive and negative lepton channels.
The event yield and background estimation of the \t-channel analysis is given in Table~\ref{tab:tchanEvent}.
The background prediction was made using a combination of data-driven and simulation based estimates.

\begin{table}[htb]
  \begin{center}
    \begin{tabular}{c|c|c}
         & $l^+$ channel  & $l^-$ channel  \\
      \hline
      \t-channel Expectation & 10.3 $\pm$ 1.8 & 4.4 $\pm$ 0.8\\
      \hline
      Background Prediction & 12.4 $\pm$ 3.3 & 8.8 $\pm$ 1.8\\
      \hline
      Event Yield & 21 & 11 \\
      \hline
    \end{tabular}
   \caption{Event yield and background estimation for the \t-channel single top-quark analysis.}
    \label{tab:tchanEvent}
  \end{center}
\end{table}

The result of the \t-channel analysis is a signal significance of 1.6 $\sigma$.
An excess over background that is consistent with \t-channel production was seen.
An upper limit of \mbox{162 pb} was placed on the cross section at the 95\% confidence level.
The systematic uncertainties in the \t-channel analysis are limited by uncertainties on the jet energy scale, \b-quark identification, and background modeling. 
Many of these systematic uncertainties are limited by statistical uncertainties in control regions and are expected to improve with the addition of more data.

Single top-quark production in the \Wt-channel gives rise to two \textit{W} bosons in the final state: one directly produced with the top quark and the other the result of the top-quark decay.
ATLAS has searched for \Wt\ production in both the single and di-lepton final states.
The single lepton \Wt\ analysis is similar to the \t-channel analysis with the additional requirement of extra jets in the event.
The di-lepton analysis requires two high \pt\ leptons, large \MeT, and exactly one reconstructed jet in the final state.
The event yields and background estimations for the \Wt-channel analysis are given in Table~\ref{tab:WtchanEvent}.
The background prediction was made using a combination of data-driven and simulation--based estimates.

\begin{table}[htb]
  \begin{center}
    \begin{tabular}{c|c|c}
         & single lepton channel  & di-lepton channel  \\
      \hline
      \Wt\ Expectation & 12.6 $\pm$ 0.9 & 2.8 $\pm$ 0.5\\
      \hline
      Background Prediction & 262.0 $\pm$ 22.8 & 12.7 $\pm$ 2.8\\
      \hline
      Event Yield & 294 & 15 \\
      \hline
    \end{tabular}
   \caption{Event yield and background estimation for the \Wt\ single top-quark analysis.}
    \label{tab:WtchanEvent}
  \end{center}
\end{table}

The analysis of the \Wt\ channel placed a combined upper limit of \mbox{158 pb} on the \Wt\ single top-quark cross section.
As in the \t-channel analysis the current systematic uncertainty is limited by uncertainties on the jet energy scale, \b-quark identification, and background modeling, which are expected to improve with the addition of more data.  

\section{Conclusion}\label{sec:conclustion}

These proceedings have presented electro-weak measurements made by the ATLAS experiment with the \mbox{45 pb$^{-1}$} of integrated luminosity collected during the 2010 data taking.
These initial Standard Model measurements have allowed ATLAS to understand its detector performance and have provided the first electro-weak physics results at 7 TeV.

\end{document}